# Stellar evolution with mass loss – comparison of numerical and semi-analytical computations.

Raul Jimenez[1,2], Uffe G. Jørgensen[2], Peter Thejll[2], and James MacDonald[3]

[1] *Nordita, Blegdamsvej 17, DK-2100 Copenhagen, Denmark*
[2] *Niels Bohr Institute, Blegdamsvej 17, DK-2100 Copenhagen, Denmark*
[3] *Dept. of Physics and Astronomy, University of Delaware, Newark, DE 19716, USA*

**ABSTRACT**
We present here results of stellar evolution calculations that include the latest advances in radiative opacities and neutrino cooling, and discuss on the basis of these models how the internal stellar structure responds to mass-loss from the stellar surface. This problem has particular importance for the development of semi-analytical algorithms for efficient calculation of synthetic stellar populations with realistic (and hence complex) mass-loss scenarios. We therefore compare our numerical results with test calculations based on a semi-analytical stellar evolution method developed by us. Although small, but important, differences between results from the two methods are revealed, the evolutionary tracks in the HR-diagram predicted by the two approaches are almost identical.

## 1 INTRODUCTION

In recent work (Jørgensen 1991; Jørgensen & Thejll 1993), a semi-analytical method was developed for analysis of stellar evolution on the red giant branch and on the asymptotic giant branch, with mass-loss included by Reimers' law (Reimers 1975) with a realistic distribution function in the mass loss efficiency parameter $\eta$. This method, which we shall refer to as synthetic stellar evolution (SSE), relies on matching high-quality observational IR data of globular clusters (GC's) red giant branch stars (Frogel, Persson & Cohen 1981) to theoretical results obtained by interpolation in grids of stellar evolution tracks. The key points of the synthetic method are that detailed stellar evolution models are used for the interpolation, and that the uncertain parameters in stellar evolution models, i.e., the mass-loss efficiency parameter in the Reimers' formula and the mixing length parameter ($\alpha$), are determined by matching the observations.

In brief, the SSE works in the following way:

1 The RGB part of evolutionary tracks in a given grid is fitted with analytic formulas which express the relation between $L$, $T_{\rm eff}$, $M$, $M_c$ (core-mass), $Z$ (metallicity), $Y$ (helium abundance) and $\alpha$.

2 The metallicity is estimated from the literature based on observed spectra which were analyzed by use of model atmospheres. The value of the helium abundance $Y$ is set to 0.24 from big bang nucleosynthesis arguments (Pagel 1992).

3 With the given values of $Z$ and $Y$ from the literature it turned out to be possible to fit all the studied GCs (Jimenez et al. 1995, Jørgensen & Thejll 1993) to the analytical expressions of point 1 by use of one value of $\alpha$. $M$ and $\eta$ were then determined by taking into account the HB morphology.

4 A very fast and accurate numerical computation of the evolution along the RGB/AGB is now performed by taking advantage of the expressions of point 1. The addition of mass to $M_c$ during a given time step in the integration along the RGB/AGB is determined on the basis of the instantaneous luminosity, the known energy generation rate, and the length of the time step. The value of $M_c$ at the end of a time step determines the new value of $L$ according to the formulas in point 1. The total stellar mass at the end of each time step is calculated as the mass at the beginning of the time step minus the mass loss rate times the length of the time step.

5 The evolution of the synthetic track is stopped when $M_c$ reaches the value determined in step 1 for the He-core flash. After this the star is evolved along the asymptotic giant branch (AGB).

The advantages of SSE over detailed evolutionary models are:

1 Determination of the $L = L(T_{\rm eff})$ relation in SSE is more accurate than in the self-consistent evolutionary tracks, because $\alpha$ is adjusted by matching to observations.

2 SSE is computationally much faster than calculation of complete models. This fact allows us to compute synthetic stellar evolution models with much more complex assumptions about, among other things, the behavior of mass-loss. We have used this advantage to predict a distribution function in the mass-loss efficiency of red giants (expressed as a star-to-star variation in Reimers' parameter $\eta$), based on fits to the horizontal branch morphology (Jørgensen & Thejll 1993).

Obviously, the SSE method does not, at a given time step, contain as detailed information about the internal stellar structure as does the self-consistent numerical solution. An important question concerning the accuracy which can be obtained by use of the SSE method is therefore how the information about the interior stellar structure is handled. Since the SSE method relies on interpolation in self-consistent models computed without mass-loss, the largest concern regards evolutionary phases with large mass-loss, i.e. the rapid evolution near the red giant tip (RGT). When comparing evolutionary tracks computed from the SSE method with tracks of self-consistent numerical models which start from the same initial conditions (i.e., with identical value of initial mass ($M_i$), $\alpha$, $Z$, and $Y$), the most critical question to ask is therefore how similar such tracks are when they end at the RGT (for example expressed in terms of the position of their RGTs in the HR diagram). One could express the most basic SSE assumption by claiming that the semi-analytical models experience instantaneous internal structure adjustment to mass-loss. Or we could say that the SSE stars have no "memory" about their past mass-loss history. Quantitative solutions to the question of the validity of this assumption depends on the ratio of the relevant time scales. Given long enough time the internal structure of the models will adjust to the total mass after mass-loss (i.e., have no hysteresis), but in particular close to the RGT, where the mass-loss rate is large and the evolution in the HR-diagram is fast, one could worry that the question of "mass-loss memory" would make it impossible to compute reliable stellar evolution tracks by use of the fast SSE method.

To illuminate this central question in the theory of synthetic stellar population computations by use of semi-analytical methods, we present here the results of computation of a number of relevant test models calculated both by a complete stellar evolution code and by use of the SSE method.

## 2 THE MODELS

We have computed a few representative stellar evolution sequences calculated using both the complete stellar evolution code and the SSE method. The metallicity is chosen to be Z=0.0002 in order to reproduce a typical globular cluster like M68. The choice of mixing length parameter, $\alpha = 1.25$ is taken from Jimenez et al. (1995).

Stellar evolution sequences, from the contracting Hayashi phase to the tip of the RGB for masses $0.75 M_\odot$ to $0.95 M_\odot$, in steps of $0.10 M_\odot$, were computed using a stellar evolution code developed by one of us (JM). Models with mass loss were calculated for values of the Reimers' parameter $\eta$= 0.00, 0.15, 0.35 and 0.70.

The stellar evolution code is a descendant of the Eggleton code (Eggleton 1971, Eggleton, Faulkner and Flannery 1973) but uses updated OPAL radiative opacities (Rogers & Iglesias 1992) for temperatures greater than 6000 K and the opacity tables of Alexander, Johnson & Rypma (1983) and Lenzuni, Chernoff & Salpeter (1991) at lower temperatures. Conductive opacities are calculated from the fit by Iben (1975) to the tables of Hubbard & Lampe (1969) for non-relativistic electrons and from the formulae of Itoh et al. (1983) for relativistic electrons. Plasma neutrino loss rates are from Haft, Raffelt & Weiss, (1994), neutrino bremsstrahlung loss rates are from Itoh & Kohyama (1983), and pair and photo-neutrino loss rates are from Beaudet, Petrosian & Salpeter (1967), modified to include the effects of neutral currents. Nuclear reaction rates are from Fowler, Caughlan & Zimmerman (1975) and Harris et al. (1983), except the triple-alpha reaction rate which is from Nomoto, Thieleman & Miyaji (1985). Coulomb interactions in the liquid phase are treated as described in Iben, Fujimoto & MacDonald (1992).

For the same parameter ranges, we have calculated SSE sequences using the method described in Jørgensen & Thejll (1993).

Table 1 compares physical parameters computed by the two methods.

In order to illustrate how physical conditions in the star are affected by mass loss, we present in figures 1 - 3 results for 3 evolutionary sequences. Sequences A and B both have initial mass $0.8 M_\odot$ but have $\eta = 0.0$ and $\eta = 0.35$, respectively. Sequence B ends with a mass of $M_{RGT} = 0.73 M_\odot$, so we have also calculated sequence C for initial mass $M_i = 0.73$ and $\eta = 0.0$. Figure 1 shows how temperature varies with density near the center of the star at the RGT. Figure 2 is the theoretical Hertzsprung-Russell Diagram near the RGT and figure 3 shows how the central temperature evolves with central density.

## 3 DISCUSSION

From figures 2 and 3, we can see how the different models evolve in the core and at the surface. At the surface the star responds to the loss of the mass, so that the sequence at $0.8 M_\odot$ with no mass-loss (sequence A) and the sequence with mass-loss (sequence B) differ increasingly as mass is lost. Significantly, the surface properties of the mass losing star come closer and closer to those of sequence C as the RGT is approached. On the other hand, the characteristics at the center of the star that is losing mass (sequence B) remain essentially the same as those of the star of the same initial mass, evolving without mass loss (sequence A). Thus, we conclude that the core is isolated from what is happening at the surface even close to the RGT. From figure 1 we can compare the internal structure of the stars at the RGT. In the core, the sequence B star has the same structure as the sequence A star, as already pointed out from simpler computations by Castellani & Castellani (1993).

Our aim, however, is to investigate the accuracy of the synthetic stellar evolution code. From the values of $T_{\mathrm{eff}}$, luminosity and core-mass at the RGT given in table 1, we see that there is good agreement between the two methods, with typical differences of the order of $0.02 M_\odot$ for the final mass, 0.02 dex for the luminosity at the RGT, $0.003 M_\odot$ for the core mass and 20 K for the temperature at the RGT. These differences are less than the error in the mathematical fit to the detailed evolutionary sequences. The 20 K difference in the temperature is smaller than the typical observational error in the temperature (± 50 K). The error in the final mass calculation ( $0.02 M_\odot$ ) is only 10 % of the spread of the mass of the stars on the HB , for this reason this discrepancy is also negligible. The discrepancy in the core mass is very small and it will produce an error in the log of the luminosity of only 0.01 dex, which is smaller than the error in the fit of the core-mass luminosity relation, and much smaller than the observational error. Therefore the discrepancies between the two methods are negligible and, therefore, both methods agree in the final values.

From table 1, we can also see that the core mass depends only weakly on total mass. If we demand an accuracy of $0.005 M_\odot$ – a value that is a good estimation due to the observational errors in $T_{\rm eff}$ and luminosity involved – then the core mass is independent of the total mass.

## 4 CONCLUSIONS

We have presented results of stellar evolution calculations that include the latest advances in radiative opacities and neutrino cooling, and analyzed the effect mass-loss has on the internal stellar structure. For globular cluster stars, our numerical internal structure calculations show that the mass-loss time-scales are such that the stellar core does not manage to readjust its structure during the rapidly increasing mass-loss rate while the star approaches its helium core flash. In contradiction to this fully numerical result, a basic assumption in semi-analytical methods is to assume that the core re-adjust its structure simultaneously with the mass-loss from the stellar surface. We have shown that in practice the external parameters (like the effective temperature and the luminosity at the RGT) are so relatively insensitive to the parameters of the core that the computationally fast semi-analytical algorithms developed by us reach the same values (within the observational obtainable accuracy) as does the full self-consistent numerical solutions, even for evolutionary phases with rapid mass-loss. This result shows that the semi-analytical algorithms are sufficiently accurate to be used to simulate stellar populations even under conditions of the strong mass-loss prevailing near the tip of the RGB in old metal-poor stellar systems.

Stars near the RGT in the globular clusters have a higher mass-loss rate than stars at most other phases of evolution, and therefore the core of stars near the RGT have a shorter time scale available to re-adjust to the changing total mass than have stars in most other phases of evolution (as is also seen from Fig. 1–3 by comparing the evolution of our models near the RGT with the evolution in the lower parts of the RGB where the mass-loss rate is smaller). Our conclusions concerning the application of the semi-analytical methods are therefore more generally applicable than just to the RGB of the globular clusters (which we have tested here). The conclusions apply to all stellar evolutionary phases with a mass-loss rate smaller than, or of the order of, the mass-loss rate of the globular cluster stars near the RGT; i.e., to most evolutionary phases of all low mass stars.

## ACKNOWLEDGMENTS

RJ acknowledges support from the EC under the Human and Capital Mobility grant 920014. JM thanks Forrest Rogers and Carlos Iglesias for making their opacity tables available in electronic form. PT thanks the Carlsberg foundation. We thank David Alexander for providing us with his new opacity tables.

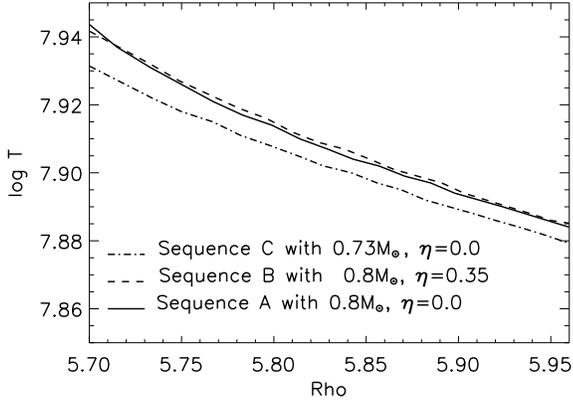

**Figure 1.** Internal structure for the models at the RGT close to the center. The plot shows the central density against the central temperature for the terminal model of three different stellar evolution tracks. Sequences A and B both have initial mass $0.8M_\odot$ but have $\eta=0.0$ and $\eta=0.35$, respectively. Sequence B ends with a mass of $M_{RGT} = 0.73M_\odot$, so we have also calculated sequence C for initial mass $M_i = 0.73$ and $\eta = 0.0$. We see how the properties of the core of the RGT models of sequences A and B resemble one another, but are markedly different from the RGT model of sequence C. This implies that the core does not feel the conditions at the outer layers.

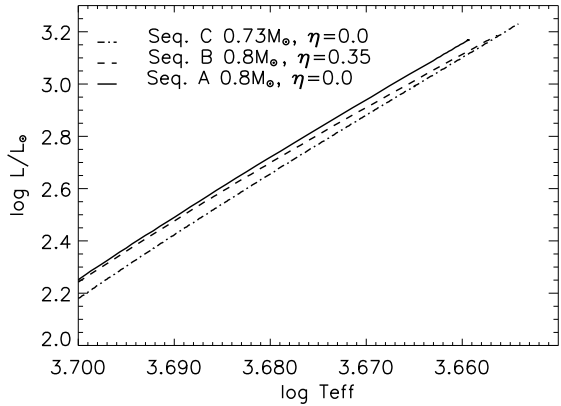

**Figure 2.** Luminosity against the effective temperature for three different tracks. This plot shows how the position in the HR–diagram do adjust to mass loss despite of the lack of sensitivity of the core to mass loss, and sequence B ends much closer to sequence C than to sequence A in this case. A, B, and C sequences with the same physical parameters as in Fig. 1.

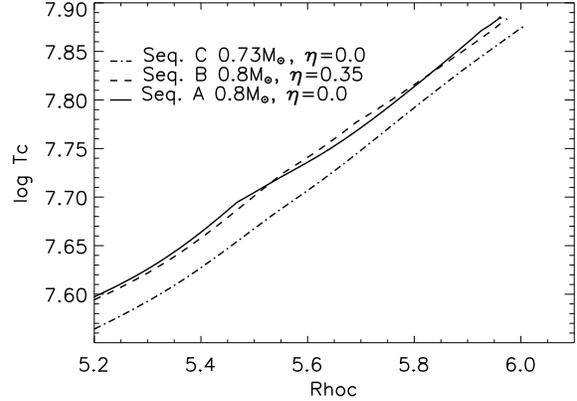

**Figure 3.** We show the evolution of the central density and the central temperature for three different tracks along the RGB. This shows how the properties of the star at the center do not change if the star suffers mass-loss. Sequences A, B, C have the same physical parameters as in Fig. 1.

**Table 1.** Properties of the sequences calculated with the detailed stellar evolution code and the synthetic code. Column (1) shows the method used to calculate the track: detailed stellar evolution code (Det.) or SSE code (Synth.), (2) the initial mass of the star in solar units, (3) the final total mass of the star at the RGT in solar units, (4) the $\eta$ mass-loss parameter, (5) the core mass at the RGT, (6) the luminosity at the RGT and (7) $T_{\rm eff}$ at the RGT.

| (1) | $M$ (2) | $M_{\rm f}$. (3) | $\eta$ (4) | $M_c^{RGT}$ (5) | $logL/L_\odot$ (6) | $T_{\rm eff}$ (7) |
| --- | --- | --- | --- | --- | --- | --- |
| Det.   | 0.95 | 0.81 | 0.70 | 0.475 | 3.18 | 4560 |
| Synth. | 0.95 | 0.83 | 0.70 | 0.478 | 3.19 | 4583 |
| Det.   | 0.95 | 0.89 | 0.35 | 0.471 | 3.15 | 4585 |
| Synth. | 0.95 | 0.90 | 0.35 | 0.476 | 3.18 | 4602 |
| Det.   | 0.95 | 0.92 | 0.15 | 0.472 | 3.16 | 4581 |
| Synth  | 0.95 | 0.93 | 0.15 | 0.476 | 3.19 | 4595 |
| Det.   | 0.95 | 0.95 | 0.00 | 0.479 | 3.19 | 4590 |
| Synth. | 0.95 | 0.95 | 0.00 | 0.476 | 3.17 | 4622 |
| Det.   | 0.85 | 0.68 | 0.70 | 0.480 | 3.20 | 4520 |
| Synth. | 0.85 | 0.71 | 0.70 | 0.481 | 3.21 | 4545 |
| Det.   | 0.85 | 0.77 | 0.35 | 0.478 | 3.19 | 4540 |
| Synth. | 0.85 | 0.79 | 0.35 | 0.479 | 3.20 | 4567 |
| Det.   | 0.85 | 0.82 | 0.15 | 0.478 | 3.18 | 4581 |
| Synth. | 0.85 | 0.82 | 0.15 | 0.478 | 3.19 | 4579 |
| Det.   | 0.85 | 0.85 | 0.00 | 0.482 | 3.21 | 4564 |
| Synth. | 0.85 | 0.85 | 0.00 | 0.478 | 3.18 | 4589 |
| Det.   | 0.75 | 0.54 | 0.70 | 0.483 | 3.20 | 4550 |
| Synth. | 0.75 | 0.58 | 0.70 | 0.485 | 3.23 | 4517 |
| Det.   | 0.75 | 0.65 | 0.35 | 0.482 | 3.21 | 4512 |
| Synth. | 0.75 | 0.68 | 0.35 | 0.482 | 3.21 | 4533 |
| Det.   | 0.75 | 0.71 | 0.15 | 0.482 | 3.21 | 4520 |
| Synth. | 0.75 | 0.72 | 0.15 | 0.481 | 3.21 | 4546 |
| Det.   | 0.75 | 0.75 | 0.00 | 0.482 | 3.20 | 4539 |
| Synth. | 0.75 | 0.75 | 0.00 | 0.480 | 3.19 | 4559 |